\documentclass[aps,prc,preprint,showpacs,nofootinbib]{revtex4}

\usepackage{epsfig}
\usepackage{graphicx,amsmath}
\newcommand\ba{\begin{eqnarray}}
\newcommand\ea{\end{eqnarray}}

\newcommand{\be}{\begin{equation}}
\newcommand{\ee}{\end{equation}}

\newcommand{\bas}{\begin{eqnarray*}}
\newcommand{\eas}{\end{eqnarray*}}

\begin{document}

\title{Excitation of physical vacuum through $\bar p p$ annihilation in selected channels }
\author{E. A. Kuraev}
\affiliation{\it JINR-BLTP, 141980 Dubna, Moscow region, Russian
Federation}
\author{E.~Tomasi-Gustafsson}
\affiliation{\it CEA,IRFU,SPhN, Saclay, 91191 Gif-sur-Yvette Cedex, France, and \\
CNRS/IN2P3, Institut de Physique Nucl\'eaire, UMR 8608, 91405 Orsay, France}

\date{\today}

\begin{abstract}
In this work we discuss the interest of measuring the ratio of of $p\bar p$ annihilation into kaon and pion pairs near the threshold region, where $S$ state dominates. It is shown that a good agreement with existing data on the ratio of the yields of charged kaon to pion pairs is obtained taking into account not only quark rearrangement in the incident hadrons, but also vacuum excitations.

\end{abstract}

\maketitle

In this paper we discuss the ratio of the yields of pion and kaon pair production in the $\bar p p $ annihilation process near the threshold region. 
 In this region, all states except $S$-states are suppressed. The main decay channel of the triplet state is to a pion charged pair, due to the absence of constituent strange quarks in the proton Fock state. On the other hand,  
the $\pi^+\pi^¯$ and $K^+K^¯$ decay channels are expected to occur with equal probability from the singlet state $^1S_0$, due to the equality of the $q\bar q$ particle-hole excitations in the physical vacuum, with $q=u,d,s$. 

The particular interest of these processes is related to the availability of high energy antiproton beams at PANDA (FAIR), in next future. The interesting kinematical region, the threshold region, can be reached in the collision of a $\bar p$ beam using the known mechanism of 'return to resonances', where the excess energy is carried by photon or jet emission.

The antiproton-proton annihilation at low energy has been studied experimentally and theoretically (for a review, see for instance \cite{Kl05}).

Strangeness production in $\bar p p$ annihilation has been considered in the literature, particularly in connection with the violation of OZI rule in vector meson production through the ratio of $\phi$ to $\omega$ yield \cite{El89,No02}, or in the polarization properties of the $\Lambda$ baryons.

At LEAR, the frequencies of $\bar p $ annihilation at rest in a gaseous $H_2$ target into pion ($f_{\pi\pi}$) and kaon ($f_{KK}$) pairs were measured \cite{Ab94}. The following result was obtained: 
that $f_{\pi\pi}=(4.26\pm 0.11)\times 10^{-3}$, $f_{KK}=(0.46\pm 0.03)\times 10^{-3}$, and their ratio: $R=f(K^+K^-)/f(\pi^+\pi^-)=0.108\pm 0.007$. 
It was also shown that the probability for $S$ state decreases when the gas pressure increases.

At threshold, where $\sqrt{s}\le 2M_p$, different states are possible, for the $\bar p p$ bound system, with quantum numbers: $^{2s+1}\!L_{\cal J}$, ${\cal J}=L+S$ equal to  $L=0 \to ^1\!S_0,~^3\!S_1..$, $L=1 \to ^3\!P_0,~^3\!P_1..$, which, in principle, may be experimentally determined.

We suggest to measure the ratios:
$$R_s=\displaystyle\frac {(p\bar p)_{{\cal J}=0}\to K\bar K}{(p\bar p)_{{\cal J}=0}\to \pi^+\pi^-},~
R_p=\displaystyle\frac {(p\bar p)_{{\cal J}=1}\to K\bar K}{(p\bar p)_{{{\cal J}=1}}\to \pi^+\pi^-}.$$
which contain useful information on the reaction mechanisms, and may give evidence for the properties of the vacuum excitations. It is expected that $R_p\ll R_S\simeq 1$. Moreover, the angular distribution of kaon production is expected to be isotropic, whereas pions would follow an angular dependence driven by the dominant contribution of triplet state, which can be observed in polarization experiments.

The mechanism considered here for $^1S_0$ production was not considered in the  previous literature. We do not consider meson production and the excited states with $L=1$, as we focus to a narrow kinematical region, where their  contribution is small.

Our approach can explain the experimental results of the ratio of pion and kaon pairs in $\bar pp$ annihilation at rest \cite{Ab94}, through the excitation of vacuum states. 

In Ref. \cite{Vo04}, similar arguments were applied to explain the $\Delta T=1/2$ rule in the decay of kaons into two pions. 

\begin{figure}
\begin{center}
\includegraphics[width=12cm]{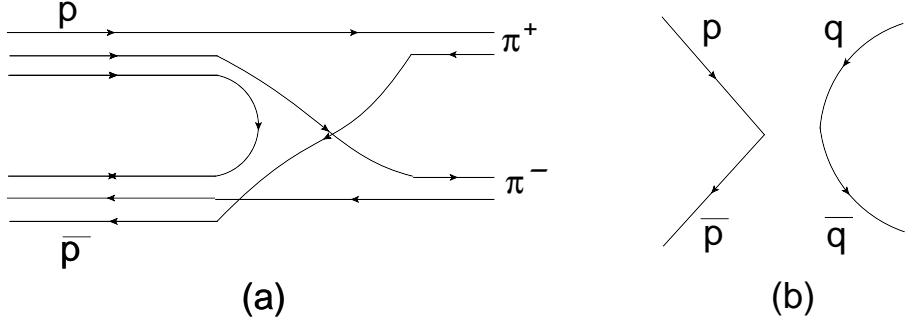}
\caption{\label{Fig:fig1} (a) Feynman diagram for the reaction $ \bar{p} +p \to \pi^++\pi^-$ through quark rearrangement; (b) scheme for production of a pair of quarks through vacuum excitation.}
\end{center}
\end{figure}

Let us focus on the threshold region, neglecting the contributions from $L=1$ states. Meson production in $\bar p $ annihilation occur through rearrangement of the constituent quarks (see Fig. \ref{Fig:fig1}a). Selection rules require that $J=1$, therefore, by this mechanism, $\pi+\pi^-$ final channel occurs from $^3S_1$ state. As the strange quark content of the proton is very small, and strange quarks come only from the sea, kaon production is forbidden through such mechanism: in triplet $S$ state only pions are produced.

However a kaon pair can be produced through a disconnected diagram Fig. \ref{Fig:fig1}b where any pair of particles can be created from excited vacuum. In this case, a singlet state, corresponding to $J=0$, $^1S_0$ can produce a pair of current quarks, which, after interacting, convert into constituent quarks and become observable as mesons. Let us underline that the probability to enhance $u\bar u$ pair of current quarks is the same as for a $s\bar s$ pair of current quarks due to the structure of excited vacuum (EV). 
The matrix element of the process $\bar p p \to \bar q q$ in singlet state, is proportional to $\bar u(p_-) v(p_+)$ (where $u,v$ are the spinors of the quark and antiquark, with four momentum $p_+$, $p_-$respectively):
\be
\left |M(\bar p p \to EV\to \bar q q)\right |^2\sim  Tr(\hat p_+-m_q)(\hat p_-+m_q)=8\beta_q^2 m_p^2,~q=u,d,s,
\ee
where $\beta_q=\sqrt{1-m^2_q/E_q^2}$,  $m_u=m_d=280$ MeV, $m_s=400$ MeV, and  $E_u=E_s=m_p$, $m_p$ is the proton mass.

The yield of pion and kaon pair production should be corrected by the phase volume ratio $\phi_{\pi}/\phi_K=\beta_{\pi}/\beta_K$.

Finally
\be 
\frac{Y_{KK}}{Y_{\pi\pi}}=\frac{1/2}{3+1/2}\frac {\beta_K}{\beta_{\pi}}
\left ( \frac {\beta_s}{\beta_{u}}\right )^2=0.108
\label{eq:eq2a}
\ee
to which we attribute an error of $5\%$ related to the constituent quark masses.
This quantity may be compared to the experimental value $R=f(K^+K^-)/f(\pi^+\pi^-)=0.108\pm 0.007$. 
The present estimation is based on the statement that pion and kaon pairs are produced 'democratically' from the vacuum excitation, whereas kaon production is forbidden in singlet $S$ state.

\begin{figure}
\begin{center}
\includegraphics[width=12cm]{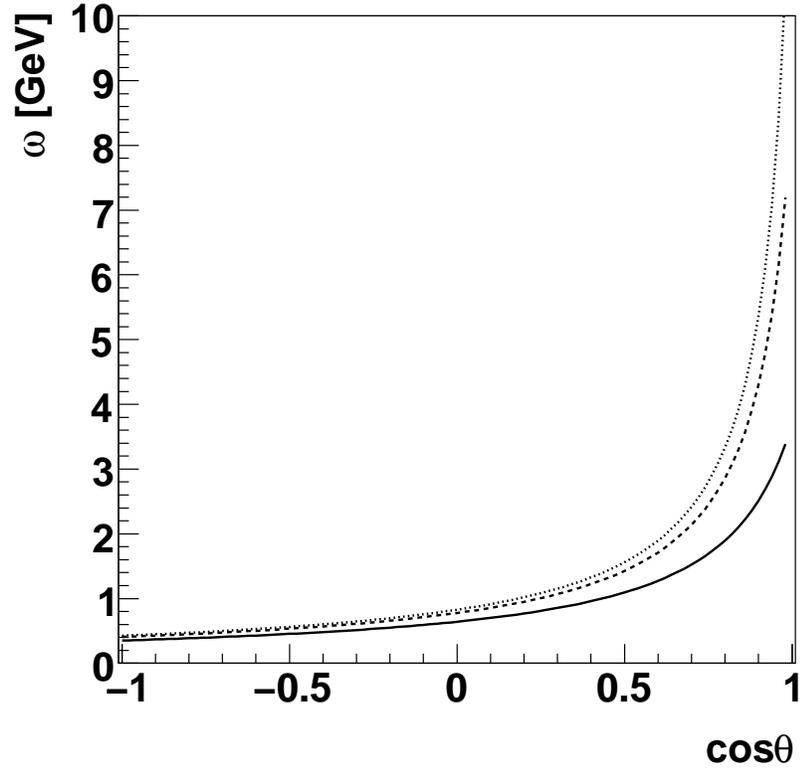}
\caption{\label{Fig:fig2}  Kinematical lines corresponding to protonium at rest in the plane $\omega$ versus $\cos\theta$, for three values of the incident energy: E=5, 10 and 15 GeV, solid, dashed and dotted lines, respectively.}

\end{center}
\end{figure}

Let us illustrate the kinematic of interest for PANDA (FAIR). The threshold region, can be reached in the collision of a $\bar p$ beam (which may have momentum up to 15 GeV) using the 'return to resonances' mechanism, where the excess energy is carried, for example, by hard real or virtual photon.
Using conservation laws and on-mass condition for a protonium ($p\bar p$ bound state, $B$ of momentum $P$):
\ba
&\bar p(p_1)+p(p_2)&\to\gamma(k)+B(P),~k^2=0,~P^2=4m_p^2,~p_1^2=p_2^2=m_p^2;
\nonumber\\
&\bar p(p_1)+p(p_2)&\to\gamma(k)+B(P),~k^2=M_X^2\gg 0
\label{eq:eqreac}
\ea
one obtains, for the case of real photon emission:
\be
(p_1 + p_2-k)^2=4m_p^2,~\omega=\displaystyle\frac{E-m_p}
{ 1+\displaystyle\frac{E}{m_p}(1-\beta\cos\theta)},
~\beta=\sqrt{1-\displaystyle\frac{m_p^2}{E^2}}
\label{eq:eqg}
\ee
where $\theta=\widehat{\vec k \vec {p_+}} $ and $\beta$  are the photon emission angle and velocity, and  $E$ is the energy  of the beam (the direction of the beam is taken as the $z-$ axis). In Fig. \ref{Fig:fig2} the kinematics corresponding to protonium at rest is shown in the plane $\omega$ versus $\cos\theta$, for three values of the incident energy: $E$=5, 10 and 15 GeV.

In the case of virtual photon emission (with subsequent conversion to jet) one has:
\be
E-\omega+\displaystyle\frac{M_X^2}{2m_p}-E(\omega-k\beta\cos\theta)\simeq m_p,~
\omega=k_0,~k=\sqrt{\omega^2-M_X^2}
\label{eq:eqgst}
\ee
where $\theta$ is the emission angle of the jet.

The probability to create a protonium state by a slow moving antiproton is finite in the limit of zero velocity, due to the bound state factor:
\be
|\Psi(0)|^2=\displaystyle\frac{\chi}{1-e^{-\chi}},~
\chi=\displaystyle\frac{2\pi\alpha}{\beta},
\label{eq:eqps}
\ee
which compensates the smallness of the phase volume of the colliding proton and antiproton. This kinematical situation is similar to the one of positronium created by slow positrons in matter \cite{Go76}.

Let us remind that the main contribution to the $p\bar p$ cross section is related to elastic $p\bar p$ scattering, which is three order of magnitude larger than the annihilation channel. For the $^1\!S_0$ state, a similar estimation can be done, when the suppression factor is calculated from the Boltzman probability: $W\sim exp(-2M/k_BT_d)$. Taking $2M=700$ MeV and the deconfinement temperature $k_B T_d=100$ MeV, one obtains $W\sim 10^{-3}$. This justifies the equal weights of singlet and triplet states, implicitly assumed in Eq. (\ref{eq:eq2a}).

A discussion of the different channels and reaction mechanisms in frame of standard QCD can be found in \cite{Do92}.

In conclusion, we gave quantitative estimation of the ratio of the yield between kaon and pion pairs, discussing two possible reaction mechanisms which differ by the quantum numbers involved - the rearrangement of constituent quarks in the reaction participants  - the excitation of the physical vacuum. 
Taking into account the mechanism of conversion of the vacuum response to $q\bar q$ state and the difference of the relevant phase volumes, a good agreement with the existing experimental data is obtained.


\end{document}